# A Neural Dynamic Model based on Activation Diffusion and a Micro-Explanation for Cognitive Operations

WEI Hui

Department of Computer Science, Laboratory of Algorithms for Cognitive Models,
Fudan University, Shanghai 200433, China

**Abstract:** The neural mechanism of memory has a very close relation with the problem of representation in artificial intelligence. In this paper a computational model was proposed to simulate the network of neurons in brain and how they process information. The model refers to morphological and electrophysiological characteristics of neural information processing, and is based on the assumption that neurons encode their firing sequence. The network structure, functions for neural encoding at different stages, the representation of stimuli in memory, and an algorithm to form a memory were presented. It also analyzed the stability and recall rate for learning and the capacity of memory. Because neural dynamic processes, one succeeding another, achieve a neuron-level and coherent form by which information is represented and processed, it may facilitate examination of various branches of Artificial Intelligence (AI), such as inference, problem solving, pattern recognition, natural language processing and learning. The processes of cognitive manipulation occurring in intelligent behavior have a consistent representation while all being modeled from the perspective of computational neuroscience. Thus, the dynamics of neurons make it possible to explain the inner mechanisms of different intelligent behaviors by a unified model of cognitive architecture at a micro-level.

## 1. Introduction: The micro-level operations of cognition

A sequence of tasks are performed in their brain when a student answers their teacher's question. These tasks are speech analysis, question understanding, knowledge retrieval, reasoning or problem solving, sentence production. On careful examination of these continuous stages, there is awareness of the sub-tasks of a problem, and the procedure to decompose the problem and infer the final decision. However, there is no consciousness of accomplishing a sub-task, and no knowledge of details that happening in the execution of a sub-task. As another example, consider the procedure for remembering the name of an acquaintance unseen for a long time. The face is in the mind, yet it's necessary to think deeply and for a long time to remember their name. It is not known in any detail how the brain associates a name with a face. As a person accomplishes each cognitive task, their brain knows what to do without being told how to do it: the implementation is veiled. This kind of "conscious blankness" is typically experienced in problem solving, perception, language understanding and production, memory and learning, and in the sensor-motor arc. The blankness, i.e. exactly which neural activities take place, needs to be explained through physiological psychology.

The realization of an intelligent behavior is essentially composed of a series of successive neural activities. Dividing the execution of an intelligent behavior into many steps means finite automata (FA) are used to model all these cognitive operations in a time sequence. This automata is activated either by outer or inner stimuli, and transforms from one state to another, finally ending with a state that stands for a directive to start a behavior, or stands for a kind of inner perception. Humans' cognitive behaviors are rich, and each cognitive operation corresponds to its own automat. At a higher level, the start of an FA might be connected to the termination of another, i.e. different FA may activate one another. In this way, different automata collaboratively perform either low-level reactions such as movement control or high-level cognition such as inference, problem solving, perception, learning, memory and speech. Each automat is a particular routine of a certain sub-task. The aim of the neural dynamic model is to describe and implement such kind of abstract automata. Related studies[1-5] are increasingly concerned with this. The spatio-temporal structure of neurons and connections, status evolution, and physical or mathematical models will explain the brain mechanism of information processing.

Memory corresponds to a local minimum of energy, or a stable neural circuit or neural dynamics. Psychology maintains that the high-level cognitive activities of humans were acquired by development. Referring to neural dynamics, this may be interpreted by how this complex structure and mechanism is established, through such procedures as conditioned reflexes and associative learning in after-birth development.

In this paper a dynamic model for activation among neurons is proposed, which is inspired by studies of neural biology and physiological psychology. Because this model explains the representation of perception and the realization of memory, it is explained how intelligence, with different types and appearances in behavior level, emerges from a general representation and implementation mechanism, and how this perception representation oriented neural dynamic model influences the core problems in artificial intelligence (AI), such as in which form the knowledge should be represented and whether a general cognitive architecture existed and should be formulated.

## 2. Model: A Network Structure by Isotropic Diffusion

Intelligent activities of different forms, including perception, association, thinking, memorization and learning as well as sensory-motion arc, share similar neural dynamics. As combinations of a variety of neurons, simple and finite, these activities, following similar sets of principles, may be described with a similar network structure.

This study was inspired by the biophysical aspect of neurons[6], particularly with respect to the following:

(1) Each neuron has a different shape. The number, length and range of the synapses and dendrites vary according to the neuron. From information processing perspective, each processing unit is not uniform, having its own specific input and output fields and being connected locally and non-uniformly. The connections are not regular, but follow statistical constraints.

(2) Each neuron has a flexible structure in terms of cell size, number of synaptic and dendric branches as well as their range. This indicates that the structure of information processing is also changeable, a major reason underlying learning.

(3) Neurons communicate with each other by sending impulses, known as action potentials. A strong stimulus larger than the threshold (typically 15 mV higher than the threshold, which ranges from -90 to -30 mV) causes neurons to send impulses of the same size and shape. Information is encoded only by the number and interval of the impulse sequences.

(4) The graded potential generated in a dendrite is linear to the stimulus stimulated in a synapse.

(5) Many local neurons exist that have no axon, and therefore no long-range connections.

They collect and propagate information locally with graded potentials on the dendrites and do not fire any action potentials.

(6) The postsynaptic potential accumulates in a spatial-temporal manner.

(7) A synapse is either excitatory or inhibitory.

(8) Excitatory and inhibitory postsynaptic potentials (EPSP/IPSP) integrate in a complex manner (especially at the axon hillock), and ultimately effect the frequency of firing an action potential.

(9) All potentials result from ions of different types flowing through the membrane. The shape of the potential reflects the summed effect of charge within a region and a period. The nature of ion flow is mathematically complex.

All of the above features contribute to a more biologically-based model for information processing, and the new model is more complex than previous artificial neural network models.

Functionally, there are three types of neurons: sensory, motion and intermediate neurons, the latter having the largest capacity. These types are modeled, respectively, as receptors, effectors and processors. According to the nature of neurons and their connections, in designing such a network structure, only intermediate neurons are considered in which the majority of information processing takes place. This is illustrated in Fig.1.

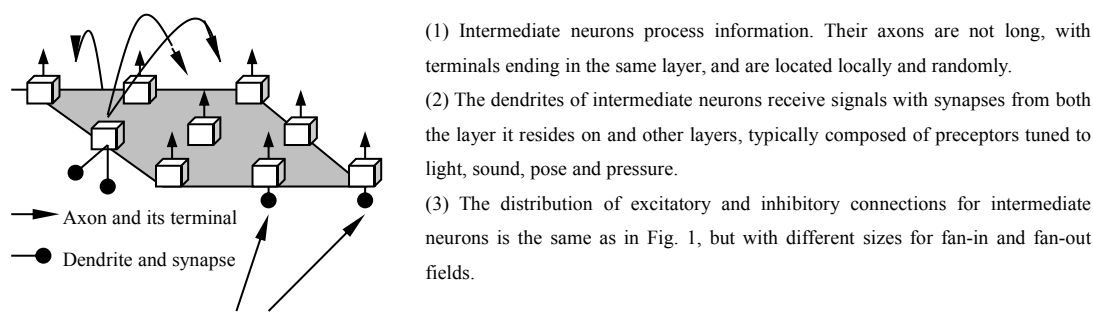

(1) Intermediate neurons process information. Their axons are not long, with terminals ending in the same layer, and are located locally and randomly.

(2) The dendrites of intermediate neurons receive signals with synapses from both the layer it resides on and other layers, typically composed of preceptors tuned to light, sound, pose and pressure.

(3) The distribution of excitatory and inhibitory connections for intermediate neurons is the same as in Fig. 1, but with different sizes for fan-in and fan-out fields.

Fig. 1 Neural Connection between Intermediate Neurons

No single neuron is ever indispensable to the whole system. Their functions compensate for each other, and a partial failure will not cause a systematic crash. From this it may be inferred that the information distributes in denser neural ensembles uniformly and redundantly. Therefore, in this model all neural connections are local, with a large number (of the scale of $10^3$) thereof, and they diffuse isotropically. The morphology of the neuron decides its connection and thus its function.

Only one layer of the hierarchical structure of the neural system, whose outputs are typically motors and inputs are generally preceptors, is shown in Fig.1. The middle level is further divided into sensing memory, working memory and long-term memory according to the three-stage memory model[7]. Of the three, working memory exhibits the dynamics of a shared structure, while long-term memory is more specific. This difference is because of the known difference of working memory (the interruption and dissipation theories) and long-term memory (the structure theory).

## 3. A Mathematical Representation of the Dynamics of Neurons

The essence of the dynamic process of the neuron system is the transmission of neuron pulses. As the signal passes through the various parts of the neural cell (the soma, axon hillock, axon, dendrite, synapse and various other membrane structures), the ions in the electrolyte move in various ways to cause the signal to undergo very complex changes (for example, the axon hillock integrates the signal). A mathematical model may be proposed for this physical and chemical process. At the molecular level, the uncontrolled and non-uniform distribution and movement of ions inside and outside the cell (such as $Na^+, K^+, Cl^-, HCO_3^-$ [6, 8]) result in the information being processed and exchanged, and they are described by using potential theory. The design in this paper is as follows:

(1) The output function of axon: When the potential exceeds the threshold, caused by ion aggregation, it triggers the active ion exchanges on the axon membrane, and the axon begins to produce action potentials. The waveform of action potential produced by a neuron is similar on the whole, even if the strength of the stimulus changes. To represent the information by means of an intermittent temporal sequence of pulses, it is assumed that the waveform of an action potential is determined by one cycle of the function $ap(t) = A|Sin(\frac{2\pi}{2w})t| = A|Sin(\frac{t}{w}\pi)|$, where $A$ is the peak of the action potential (approximately 80mV), $w$ is the wave width (approximately 1 msec), and time $t \in [0,1]$. Let the resting potential of a neuron be $rp$, then its threshold potential $Th$ is approximately $rp+15$mV. From time $t_1$ to the beginning of the next stimulus which causes an axon response, the output function of the axon is the activating degree of neuron $Active^{(n)}(t_1)$ at time $t_1$ (this is defined in Section 4), which is related to time $t$. Considering the summation efficiency in time, it may be induced that

$Efferent(Active^{(n)}(t_1), t) = Efferent(t_1) +$

$$\begin{cases} rp, \text{When } Active^{(n)}(t_1) = 0 \\ rp + \frac{Active^{(n)}(t_1)}{k}|Sin(\frac{t}{w/k}\pi)|, \text{When } 0 < Active^{(n)}(t_1) \leq Th^{(n)}, k > 1 \\ rp + m \times ap(t), \text{When } Active^{(n)}(t_1) > Th^{(n)} \end{cases} \quad ①.$$

The characteristics are chosen, such that the half cycle of the sine function is used from an arbitrary point increasing to its maximum, then decreasing until its value is zero. Coefficient $k$ is the rate of compression; while $m$ is the number of action potentials produced. For the calculation let $\Delta = Active^{(n)} - Th^{(n)}$, and use the bounded function $m = \left\lceil \frac{e^\Delta - e^{-\Delta}}{e^\Delta + e^{-\Delta}} / \beta^{(n)} \right\rceil$, where $\beta^{(n)}$ is a representation of the constant, transmitting-pulse limit of a neuron. Due to the minimum interval between the opening and closing of the axon ion's channel, the frequency of pulses sent by neurons is finite.

(2) The function of the post-synaptic terminal (input function): The post-synaptic terminal changes the permeability of the cell membrane and results in the ion exchange and potential change, when the pre-synaptic terminals of the axon terminals of other neurons release neurotransmitters. This is equivalent to the input of a neuron, with its magnitude proportional to the stimulus strength. For this reason, the input function corresponding to a certain post-synaptic terminal *i* on a certain neuron *n* may be defined as $Afferent_{(i)}^{(n)}(t) = F(Stimulus(t))$, where *F(x)* is the function of homomorphic and conformal transformation against the stimulus strength, and is optional. *Stimulus(t)* is the temporal sequence of action potentials, which comes from the pre-synaptic terminals (when the stimulus is strong), or it is the temporal sequence of a negligible disturbance (when the stimulus is weak). In this paper, the following function is applied:

$$Afferent_{(i)}^{(n)}(t) = C_{(i)}^{(n)}(t) \int_{t-T}^{t} (Efferent_{(i)}^{(pre)}(t) - rp^{(pre)})dt \times \frac{d(Efferent_{(i)}^{(pre)}(t))}{dt} \qquad ②,$$

where *T* is the time interval between the potential effect of the pre-synaptic terminal being applied and fading away, and *pre* denotes the pre-synaptic terminal. The function ② makes the potential difference depolarized, which means that the accumulation of the change of signal transmitted by the pre-synaptic terminal is transformed by a unique characteristic function $C_{(i)}^{(n)}(t)$ corresponding to the post-synaptic terminal, that records the changing trend of signals from the pre-synaptic terminal. The above represents the case of the excitatory postsynaptic potential (EPSP). In the case of the inhibitory postsynaptic potential (IPSP), the function is given by

$$Afferent_{(i)}^{(n)}(t) = -C_{(i)}^{(n)}(t) \int_{t-T}^{t} (Efferent_{(i)}^{(pre)}(t) - rp^{(pre)})dt \times \frac{d(Efferent_{(i)}^{(pre)}(t))}{dt} \qquad ③,$$

and the potential difference is hyperpolarized.

(3) The integration of input function: The input signals may come from different synapses (spatial summation efficiency), or from synapses at different moments (temporal summation efficiency). Signals may change when they are transmitted inside neurons or on the cell membranes of neurons. For example, a signal decreases as it progresses. Many signals would be integrated unknowingly (such as a very complex superposition of signals). As signals are aggregated on the axon hillock this results in an integrated efficiency and affects the output signal of the axon. This implies that the shape of a neuron has a concrete affect on its information processing. For example, the synapses distant from the axon terminals are more susceptible to the changes of input signal than the synapses nearer to the axon terminals. This is one of the characteristics in which electrochemical signals differ from electronic signals. Therefore, the accumulation of signals are expressible as $Afferent^{(n)}(t) = \sum_{i=1}^{S} H_i(Afferent_{(i)}^{(n)}(t))$, where *s* is the total number of synapses on dendrites and on the cell body of neuron *n*. The function ④ is selected, which is attenuated on the strength of the signal and fits the transmission changes produced by the input signal from synapses.

$$H_i(Afferent_{(i)}^{(n)}(t)) = \frac{Afferent_{(i)}^{(n)}(t)}{K_i(t)}, K_i(t) = k_i t + b_i, \ k_i > 0, b_i > 1 \qquad ④.$$

(4) The activation function of a neuron, which is a Sigmoid-type function, is

$$Active^{(n)}(t) = c^{(n)} \times \frac{1 - e^{-k^{(n)} Afferent^{(n)}(t)}}{1 + e^{-k^{(n)} Afferent^{(n)}(t)}} \quad , \text{ where } c^{(n)} \text{ and } k^{(n)} \text{ are respectively the}$$

characteristic constants to control the scope of its range and the strength of its step.

Figure 2 illustrates the different positions at which the four functions above are calculated.

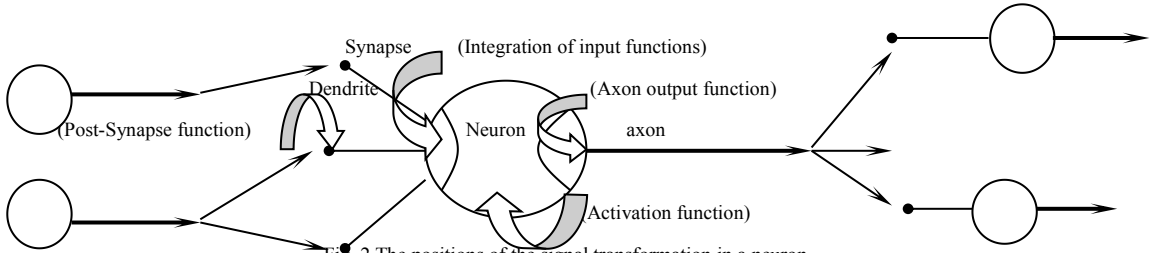

Fig. 2 The positions of the signal transformation in a neuron

Above is a mathematical description of the actions of a neuron's dynamics. This is inadequate because the mental activities which correspond to cognition or intelligent behavior are governed jointly by a mass of active neurons in a kind of collective effect. When a large number of neurons activate one another in a systematic way, the interactions among them become very complex. The assembled dynamics of a neural network are discussed in the next section.

## 4. Diffusion of Activation and a Mathematical Representation for Multiple Neurons' Assembled Dynamics

When a stimulus from outside appears at a certain time, whether it is maintained in the same model or during this period it changes continuously or discretely, it is measured by a temporal signal function which activates a mass of sensory neurons at the same time or one after another which diffuse towards the middle neurons in an instant. When mutual affection between neurons occurs, an evolving dynamic process begins. The activation transmission occurs by means of diffusion, and the direction of transmission is dependent on the gradient of potential along the distribution of neural connections. This not only ensures the uniformity of transmission, but also makes information represented symmetrically and distributedly. Transmission in this way is consistent with the physiological explanation of memory. Diffusing activation in this fashion is another explanation for the dynamic feature of inter-neuron signal processing. The dynamic process is described as follows:

(1) At time $t$, for all nodes in the middle neurons, their degrees of activity are the set of special states: $S_{(t)} = \{Afferent^{(i)}(t), i \in N\}$, which is viewed as an instantaneous distribution or a snapshot of oscillating strength.

(2) A continuous stimulus from outside maps onto the network's middle layer neurons, which are activated. When discretizing this continuous process for short periods, the stimulus assigns the middle neurons with a time sequence $S^s$ of states.

(3) If the transformation is synchronized with time for all neurons in the entire network, the process of information is a function of $T(S_{(t)}, S^s)$.

(4) The function itself is the reflection of the network structure, so an aspect of it, such as the parameters or connections, would be changed by the transmission, and as a result, the function would be changed into a new one: $T'(S_{(t)}, S^s)$.

Thus, the whole dynamic process begins with $S_{(0)}$, and the set $\{S^s_{(t)}, t=1,2,\cdots\cdots,n\}$

corresponds to a state changing with time. This is described by the following formulas (in which ★ is a transforming operator, by which the function $T$ is transformed by the results of the transformation. The algorithm to achieve this, appears later.)

$$S_{(1)} = T(S_{(0)}, S^s_{(1)}), \quad T^1 <= T \bigstar S_{(1)},$$

$$S_{(2)} = T^1(S_{(1)}, S^s_{(2)}), \quad T^2 <= T^1 \bigstar S_{(2)},$$

…… ……,

$$S_{(n)} = T^{n-1}(S_{(n-1)}, S^s_{(n)}), \quad T^n <= T^{n-1} \bigstar S_{(n)},$$

$$S_{(n+1)} = T^n(S_{(n)}, 0), \quad T^{n+1} <= T^n \bigstar S_{(n+1)},$$

…… ……,

$$S_{(n+k)} = T^{n+k-1}(S_{(n+k-1)}, 0) = 0, \quad T^{n+k} <= T^{n+k-1} \bigstar 0.$$

The steps in the earlier $n$ iterations are the activating phase, while in the latter $k$ iterations, they are the stabilizing phase, during which the network's activation is gradually attenuated and the network becomes calm. The final result of the dynamic process permanently changes the network structure and the parameters therein; in other words, this is the result of learning.

The above evolution of states is a passive process. It is analogous to when a fire engine is seen for the first time, and someone cautions to give way to it. Then, in the future, whenever the whistle of fire engines is heard, drivers will pull over to the side of the road. This is generally the learning process in humans: Others teach how to deal with something encountered for the first time, and thereafter it is no longer necessary to be told what to do and how to do it. Knowing how to deal with future similar situation is a significant aim of learning. Such recurring processes are active reaction. Therefore, psychologically, reflexes are established, in which the realization of learning is reduced to a micro-level. That is, the induced dynamic process is produced in a linked manner, so that even if there is only a small stimulus next time, the system is able to repeat the entire previous process. For example, although the set $\{S^s_{(t)}, t=1,2, \cdots\cdots, m(<n)\}$ is not entirely based on the process mentioned above, the reaction traces of the dynamic process continue to exist, and the whole process is able to be awakened and continued.

Thus the key to the learning algorithm is to keep the traces of neurons connecting in the network, i.e., to make the states of neurons evolved or triggered successively and orderly. The typical content required for learning is to combine the perception of stimulus (input) with the proper respondence (output). In the final analysis, actions occurring in the synapses are significant to learning. Neurobiological research indicates that long-term memory needs (a) the involvement of the hippocampus; (b) a change in the efficiency of synapse transmitting signals; (c) the representation and translation of genes; and (d) the formation of new synapses. Learning durably changes the neurons, but the significance of these results for information processing remains unclear.

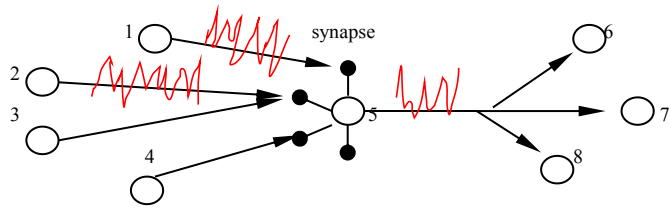

(a) The signal transmits by way of broadcast to the synapses; the effects of this signal imposing on other neurons are determined by the neurons themselves. (b) The action potential sequence implies encoding of information. (c) The essence of learning connectivity is to obtain and maintain stable temporal relations of neurons with an orderly activity, and establish mutually distinguishable circuits of activation.

Fig. 3. Connection-learning in neurons

In general, sensory neurons are activated by an outside stimulus, and they have two types: semantic-enabled activation and oscillating-derived activation. Learning occurs between the two groups of neurons in the first class. Because connections are not naturally implicated in the first case, circuits under this case are not compulsory, and therefore are established by force. Figure 3 illustrates a generic connection model, for which several possible situations are listed and connection-learning is discussed:

(1) If neuron 5 is active while others around it are inactive, it constantly radiates an action potential sequence. (It may be inferred that neuron 5 is activated by a sensory neuron, and as long as the stimulus remains consistent, this kind of radiation continues). Furthermore, various neurons in the set {6,7,8} may be activated because of the effect of time summation, which causes paths which are linked to radiate neuron 5's information.

(2) If neuron 5 is active, and only one side of set {1,2,3,4} or {6,7,8} exists as an active node, then the neurons in a receiving state undergo modulation learning (see below).

(3) If neuron 5 is active, and there are active neurons both in set {1,2,3,4} and in {6,7,8}. Neuron 5 and the neurons in set {6,7,8} undergo modulation learning together.

(4) If neuron 5 is inactive, but there are active neurons in set {1,2,3,4}, then neuron 5 receives action potentials, and due to the effects of space and time summation it may be activated.

In addition to the previous discussion, excitatory synapses may exist between two active neurons; this is the case which is usually studied. On the contrary, if these synapses are inhibitory, the post-neuron may be restrained; it is then necessary to apply more neurons besides these to calculate. Consequently, the information is dispersedly and safely represented.

Modulation learning occurs in the neurons which are located at the receiving end; the aim is to record the correlating model of activating neurons. Hebb's law states that the relations between two simultaneously active neurons must be strengthened; however, it's necessary to propose an appropriate way to make this feasible. This study's learning algorithm is similar to Hebb's law in principle. This algorithm does not apply a typical method for changing the weights of connections, and a new concept of differentiability is adopted. Neglecting this would result in an activation flood, and then the network would lose its representation purpose. In the neuron-level model of cognition, the neurons are the sole and final decision-makers; there is no other monitor at a higher level. Representation differentiability ultimately relies on differentiability of activation for each neuron.

The signal coming from the sensory neurons will activate several middle neurons, each of which will activate other middle neurons by way of diffusion. From the viewpoint of a directed graph, several maximal soaking branches will be structured, and the target of learning is to connect all these branches together. For example, in Fig. 5 the active nodes are 1, 2, 5 and 6 and the first two nodes 1 and 2 are connected in the same branch, but separated by the branch between 5 and 6. (The neurons that activate these four nodes are omitted here). The problem is how to connect these four nodes when nodes 1 and 2 are next activated, their outputs will activate node 5, then node 5's output will activate node 6, i.e., a new stimulus source is added to the network. From a microscopic view, the potential in node 5 which is produced by the activation of nodes 1 and 2 must exceed its own threshold. Likewise, the potential in node 6 produced by the activation transmission of node 5 must exceed the threshold of node 6. If {3,4},{5},{7,8} are in a similar situation, but correspond to another memory, as neuron 5 is shared for representation, it must be capable of distinguishing these two different reaction chains. Thus it must be asked if these are a new activation from {1,2} or from {3,4}, or some other novel combination.

Active nodes 1 and 2 transmit action potential pulses. Such waves are regarded as characteristic waves for representing the current information. Each post-synaptic terminal will do a different transformation on its input pulse sequence, e.g., in this algorithm, for each post-synaptic terminal, one function $C_{(i)}^{(n)}(t)$ is selected from a set of sine functions {$Sin(nt)$, $n \in N$}, and the function $\frac{d(Efferent_{(i)}^{(pre)}(t))}{dt}$ is replaced by $\omega_{(i)}^{(pre)}(t)$ (the rate of the action potential frequency at the pre-synaptic terminal to time). Thus, the modulation function of the input information is the following function: $Sin(n_k t) \times \sum_{i=1}^{m_k}[\omega_{(k)}^{(pre)}(t)(t_i - t_{i-1})] \times \omega_{(k)}^{(pre)}(t)$, i.e., the frequency of the output pulses sent by the pre-synaptic terminal is used as the amplitude-modulation of the input function of the post-synaptic terminal. $T$ is partitioned into $m_k$ periods, ($t_0, t_1, \cdots, t_{m_k}$) for the purpose of including the summation efficiency of time. The potential for neuron 5 to represent current information is

$$\sum_{k \in \{1,2\}} \{Sin(n_k t) \times \sum_{i=1}^{m_k}[\omega_{(k)}^{(pre)}(t)(t_i - t_{i-1})] \times \omega_{(k)}^{(pre)}(t)\}$$, which is a composite characteristic wave.

At the same time the output of active neuron 5 itself is also a particular respondence $Efferent(t)$ with respect to its input. The learning algorithm which is found makes $Efferent(t)$ the dependent variable of a mapping $F_{(5)}$,

$$F_{(5)}: \sum_{k \in \{1,2\}} \{Sin(n_k t) \times \sum_{i=1}^{m_k}[\omega_{(k)}^{(pre)}(t)(t_i - t_{i-1})] \times \omega_{(k)}^{(pre)}(t)\} \xrightarrow{F_{(5)}} Efferent(t).$$

Information is thus represented through neuron 5, such that additional mapping pairs like these are found, and $F_{(5)}$ becomes reasonably functional.

Because the issue of concern in practice is the number of pulses and the measurement of their sparseness (or denseness), the following simplification is made: measuring the characteristic wave → (mapping) → calculating the average density of the output pulses. The key to mapping is the measurement of the characteristic wave. To achieve a "distinguishable" target, a measurement is utilized that calculates the curve integral of a changing force moving along a curve. As such, it not only reflects the significance of the shape of curve, but also makes different works unequal, because the same force might move along different curves (waves). It is assumed that the force is $Active^{(n)}(t)$, whose direction is the same as the tangent direction of the curve, that applies along the characteristic wave (regarded as C), and whose work is equal to

$$E = \int_C Active^{(n)}(t)ds$$

$$= \int_{T_s}^{T_e} Active^{(n)}(t)\sqrt{t^2 + (\sum_{k \in \{1,2\}} \{Sin(n_k t) \times \sum_{i=1}^{m_k} [\omega_{(k)}^{(pre)}(t)(t_i - t_{i-1})] \times \omega_{(k)}^{(pre)}(t)\})^2} \, dt,$$

where $T_s$ and $T_e$ are respectively the beginning and ending points in the active period of neuron 5,. Both this $E$ and the frequency $\omega$ of output pulse are written as a sequential pair $(E, \omega)$. When such sequential pairs are continually cumulating, $F_{(5)}$ is simplified to a fitting function $\omega = F(E)$. Thus, the learning algorithm is an uninterrupted process, in which all sequential pairs are collected and the fitting function is modified by those newly sequential pairs during the time neuron 5 remains active.

With the accumulation of memory experience, the set $\{(E_i, \omega_i)\}$ continuously expands and $|\{(E_i, \omega_i)\}|=n$. Then the average fitting algorithm of a parabola is applied (It is reasonable to use other curve smoothing algorithms). In this paper, the polynomial $\omega = a_0 + a_1 E + a_2 E^2 + ... + a_n E^n$ to fit $\omega = F(E)$ is applied, where the $a_i$s are determined by the following simple equation group.

$$\begin{cases} \frac{1}{n}\sum_{i=1}^{n}\omega_i = a_0 + a_1(\frac{1}{n}\sum_{i=1}^{n}E_i) + a_2(\frac{1}{n}\sum_{i=1}^{n}E_i^2) + ... + a_n(\frac{1}{n}\sum_{i=1}^{n}E_i^n) \\ \frac{1}{n-1}\sum_{i=1}^{n-1}\omega_i = a_0 + a_1(\frac{1}{n-1}\sum_{i=1}^{n-1}E_i) + a_2(\frac{1}{n-1}\sum_{i=1}^{n-1}E_i^2) + ... + a_n(\frac{1}{n-1}\sum_{i=1}^{n-1}E_i^n) \\ \frac{1}{n-2}\sum_{i=1}^{n-2}\omega_i = a_0 + a_1(\frac{1}{n-2}\sum_{i=1}^{n-2}E_i) + a_2(\frac{1}{n-2}\sum_{i=1}^{n-2}E_i^2) + ... + a_n(\frac{1}{n-2}\sum_{i=1}^{n-2}E_i^n) \\ \vdots \\ \omega_1 + \omega_2 = 2a_0 + a_1(E_1 + E_2) + a_2(E_1^2 + E_2^2) + ... + a_n(E_1^n + E_2^n) \\ \omega_1 = a_0 + a_1 E_1 + a_2 E_1^2 + ... + a_n E_1^n \end{cases},$$

Finally, the fitting function $\omega = F(E)$ is achieved.

## 5. Memory Encoding Algorithm based on Structure Characteristics

Different neurons emit different firing sequences under the different stimuli outside and/or inside, as shown in Fig. 5. Although this is not complete for all forms of memory, it indicates that the neuron sequence of action potentials could be a representation or a code to certain information.

In order to remember stimuli with a clear meaning, brains ultimately produce a physical representation of them. Their form is an evolutionary process dominated by a group of dynamic neurons which is finally stabilized. The collective dynamics is met by temporal orders, quantitative constraints, and is solidified by a distributed micro-structure. Nevertheless, a wide adaptable range will ensure that all kinds of information are memorizable and recalled reliably by brains, and even the perceptions, which are caused by similar stimuli onto different neural systems, will not be much different. In this paper, a memory of something has been formalized into a unique connected structure, together with the processes of activation transmitting and state stabilizing. The structure is regarded as a subgraph of the global directed graph. The pattern of memory, which is realized by a network with random and localized connections, has two distinguishing methods: in terms of mode of unique structure, in terms of mode of unique encoding. It ultimately achieves discriminability for different contents to be remembered.

After the above discussion, it is clear that the neuron structure is sufficiently important that it is necessary to include its construction in the memory coding algorithm. The construction algorithm for this is given below:

*The first step: constructing a neuron network*

1a. For use with neurons in the same layer of a network which are homogeneous, to produce a group of neurons with a certain number scale, and to structure an array with a certain physical thickness (concerning implementation through hardware). This leads neurons to be distributed randomly and evenly within the network. The purpose of this is that dendrites and axons are able to gain enough space to extend, and then form mutual connections among neurons that are responsible for processing the collecting information.

1b. For each neuron, to produce random connections to others within a local ball-like area, and including two kinds of synapses. The first kind includes all connections between its own dendrites and other axon terminals, which take charge of collecting the input information. The range (referring to other neurons) of collecting information is known as its input range. The second kind includes all connections from its own axon terminals to other neurons' dendrites, which take charge of outputting information, and its scope of radiating information is known as its output range. In this neuron array, a number of input connections come from sensory neurons, and also a number of output connections go towards effector neurons. It is the selective connections between middle neurons and sensory neurons, and their proper response to the strength and nature of stimulus received from sensory neurons which makes the middle neurons take a role in the feature representation.

1c. To set the value of each parameter produced by each neuron, including the amplitude and width of threshold potential, rest potential, action potential, and $\beta^{(n)}$, to set the value of each parameter in the axon output function, post-synaptic terminal function, input integrating function and activation function.

*The second step: memorizing through a unique structure.*

2a. If a stimulus appears outside, first of all, let selected sensory neurons be activated, and then to activate middle neurons in the middle neuron array constructed in the first step.

2b. Let the middle neurons process and memorize information, those activated middle neurons send the unique firing sequence out, then diffuse and transmit along the connections of neurons, and produce two types of connections:

(1) First, in the stage of uncontrolled radiation: The action potentials sequence diffuses and transmits along inherent connections. Along the path the sequence is transformed by an incoming neuron's post-synaptic terminal function, input integrating function and activation function. Consequently new middle neurons might be activated, and these newly awakened middle nodes produce their own firing sequences, based upon their own axon output functions, and thus continue further transforming the information. Repeating this process will awaken a great number of middle neurons which are not directly stimulated by the active sensory neurons. This achieves the purpose of representing information redundantly and distributedly. These activation relations keep a kind of specialized neural circuit in which neurons are sensitive to particular stimulation.

(2) Second, in the semantic linking stage: The neuron circuits that were built in the first stage might be isolated, so that is inadequate for representing the whole stimulus. Thus, a further need is to link physically the natural active circuits together in as many directions as possible. This is the modulation learning process shown in Fig. 5. For the purpose of gaining a highly reliable representation, to solve the fitting function $\omega = F(E)$ and thus to realize the semantic association (a detailed explanation is given in Section 4), it finally makes a local natural circuit gain many means of being activated. At the same time, with the existence of redundancy at neurons and their connections, it will also lead to a highly realizable semantic association.

(3) Finally, merging two kinds of neuron circuits mentioned above into a pulse-transmitting circuit and consolidating the circuit will realize a memory of the current stimulus. This memory schema is signed by the unique active neurons group and the unique connections within the group, and the action potential pluses transmitting under this connection model are totally different from each other.

2c. An event is divided into multiple fragments of successive scenes, enabling the memorization each of fragment, subsequently joining these fragments together. Working memory keeps information for a short time. During the content shift in working memory, there exists a non-empty intersection between the active traces of the neurons corresponding to the previous fragment, and the traces of the newly activated neurons corresponding to the next fragments. That intersection includes simultaneously active neurons irrespective of space or time. Before the memory of the preceding fragment totally fades away, nodes belonging to the preceding and successive fragment activate each other and even correlate, for the purpose of remembering an event whose fragments are correlated as a time sequence. That event is then recalled through only partial fragments. The causal associations were memorized through learning like this, and under similar routine of neuron reaction it's possible to recur. A familiar example is learning and applying the "sense-motor" model. After studying traffic rules, a person would actively and consciously start their neural reactions for decelerating and braking when they see a red light at an intersection, and take the correct actions by commanding the muscles in their hands and feet.

A deeper purpose of studying memory is to understand the model of learning and acting, and how human memory makes intelligence or thinking possible and drive them. The algorithm in 2c has wide significance for cognitive representation.

## 6. Network Performance Analysis
6.1 Stability analysis

That activation transmitting may be steady for a limited time, after outside stimuli disappear is the stability and convergence problem, which is the prerequisite for an item finally to form a stable inner representation of itself. Several factors are related to the above problem: (a) There exists a minimum time interval between the two activations of each neuron. Within this period, the ions on both sides of the membrane will re-aggregate and maintain certain potential differences, and the neuron is no longer be active. Thus, in this paper the neurons are assumed to have the above character, and for all neurons, the intervals are nearly the same and these intervals are longer than their active periods. (b) The axon output function and neuron activation function are bounded. (c) The input function is gradually fading. (d) A threshold exists for each neuron that limits the neuron activity. (e) There exist equivalent inhibitory synapses to the excitatory synapses. As the memory network is a complex nonlinear system, quantitative analysis is another complex task [9, 10, 11]. In this paper, the following analytical methods are adopted, and the quantitative constraints to meet neurons working normally are considered.

The following assumptions are made. There are $N$ ($N>0$) neurons in the network, and for each neuron, its axon terminals and dendrites are evenly distributed in its own neighboring region. The number of input/output connections decided individually by the axon terminals and dendrites is $M$ ($>0$). The mathematical expectation of the number of inputs which are needed to activate a neuron is $d$. $S$ ($>0$) neurons are activated at time $t$ ($t=0$), if the stimuli appear, and these activated neuron sets are recorded as $\{S_i \mid Active^{(S_i)}(0) > Th^{(S_i)}\}$. The next step is to study the probability that a neuron in the remaining $N$-$S$ neurons is located in the range of outputs of $S$ active neurons. It is clear, however, that the trend is for the probability to be close to 1, when $S$ and $M$ are increasing.

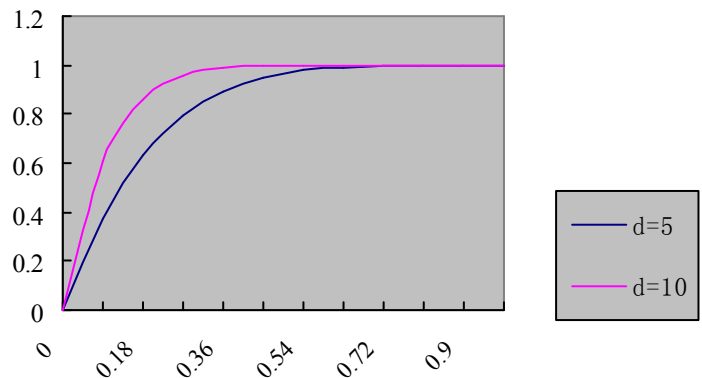

Fig. 4 Plots of function $q = 1 - (1-p)^d$

For this purpose, a power function was selected as the approximate probability function, i.e., $p = (\frac{S}{N})^{\frac{1}{M}}$, $0 \leq \frac{S}{N} \leq 1$ and $0 < \frac{1}{M} < 1$. Thus the probability of a neuron in the remaining $N$-$S$ neurons being activated, is $q = 1 - (1-p)^d$ (see Fig.4). The mathematical expectation of the number of neurons activated by $S$ active neurons in the remaining $N$-$S$ neurons is $(N-S) * q$ (this is similar to the problem of "calculating the number of missiles hitting the target when firing $N$ missiles, if a single missile hit probability is *q*"). All neurons have some active cycle, during which neurons originally resting may be increasingly activated, but due to the existence of the foregoing restriction in (a) the total of the active neurons will reach saturation point after some time. Then the total number of active neurons will gradually decline. This trend is shown in Fig. 5; with *t* growth, the number of active neurons reaches a maximum at the point *Tmax*, then the number begins to decrease because of the deadlines approaching for the active neurons in set *S*.

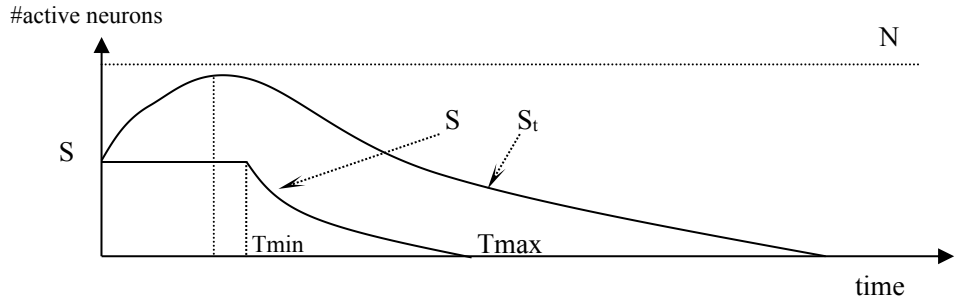

Fig. 5 Trend of number of active neurons after a single stimulus

The following recursive formula expresses this process of change in the number.

When *t*=0, $S_0 = S$;

When *t*=1, $S_1 = S_0 + (N - S_0) \times q_1$, $q_1$ is the *q* mentioned before;

When *t*=2, $S_2 = S_1 + (N - S_1) \times q_2$, $q_2$ is acquired based on $S_1$ according to the method for *q*;

……

When *t*=i, $S_i$

When *t*=i+1, $S_{i+1} = S_i + (N - S_i) \times q_i$.

Using time *t* as the variable, the general formula is obtained by substituting and spreading the above formula one item at a time:

$$S_t = S\prod_{i=1}^{t}(1-q_i) + N(\sum_{i=1}^{t}q_i - \sum_{i \neq j}^{C_t^2}q_i q_j + \sum_{i \neq j \neq k}^{C_t^3}q_i q_j q_k + \cdots + (-1)^{k+1}\sum_{j}^{C_t^k}\prod_{m=1}^{k}q_{j_m} + \cdots + (-1)^{t+1}\prod_{i=1}^{t}q_i)$$

Note that $\forall q_i < 1$, therefore, $S_t$ does not expand rapidly with *t* growth; but to force $S_t$ to decline rapidly after a time *t* (>*Tmin*), the following conditions must be met: (a) $S$ must decline rapidly; (b) *q* must have a rapid decline together with a decline of $S$; (c) *d* and *Th* must be large enough; (d) in practical applications, the number of elements in set $S_t \cap S_{t+1}$ are less than the number of elements in set $S_t$. This means that the number of newly active neurons does not increase too much every time. It is not difficult to meet the five conditions mentioned above and the input and output uniform distribution (this guarantees that an average of half the activating actions takes place among neurons which are already active).

The above analysis yields a set of parameters which guarantees that the network returns to a calm state a short time after activating the network.

6.2 The rate of successfully recalling

The extent that a sample is obscure, but it correctly recalled depends on the complexity of the sample. It is inversely proportional to the complexity and the fuzzy degree of the sample. This is experimental inference requiring further investigation.

For the problem that neural networks recover a defaced sample through memories, the succeeding rate involves many factors in a model based on structure learning. These include the complexity of the original sample, the uniformity of the sample's representation in the memory network, the defaced situation of the test sample, the distribution of abnormal changes in the test sample, the uniformity of the distribution of the test sample against original samples, the similarity of the test sample itself and its abnormal change to other samples, the sensitivity of the above similarity of other samples to the original sample, the existence of bridge nodes or seed nodes in the representation of the original sample and the values of their threshold. Because the model is based on random connections and the parameters have different settings, and the consolidation process is a process of nonlinear dynamics, for all of the above factors, it is worthwhile considering other research, especially for factors No.2, 6, 7, 8, and 9. Therefore, in this paper an analysis is made in the most basic situation where any one neuron connection meets the binomial distribution $B(n, p)$ or normal distribution $N(\mu, \sigma^2)$ in the global scope, but shows uniform distribution in its own neighborhood. There are two cases that lead to memory failure. The first is when a memory of the sample cannot be formed. The second is when the test sample is substantially different from the original sample.

In the first case, if a sample cannot generate a fully-connected representation, it means the sample might be expressed on several isolated maximum-connected branches. Consider the probability of forming two isolated maximum-connected branches by $S$ active neurons. Let those branches divide $S$ into two parts $x$ and $y$, so that $|x|+|y|=|S|$. If the output closure of $x$ cannot cover $y$, but the input of $x$ covers $y$. $y$ is used as the output covering set, $x$ will be expanded into the maximum branches by $y$, and in this case, the output coverage only needs to be considered. Let the output closure formed by $x$ and $y$ be the two sets $S_x$ and $S_y$, where $S_x \cap S_y = \emptyset$, $\forall n(n \in S_x \wedge n \notin S_y)$, $\forall n(n \in S_y \wedge n \notin S_x)$. Let the probability be $p_1 (0 < p_1 < 1)$, that at least one of the outputs that produced any one neuron in $S_x$ is located in $S_y$. Therefore, the probability of $\forall n(n \in S_x \wedge n \notin S_y)$ is $(1-p_1)^{|S_x|}$. Similarly, if the probability is $p_2 (0 < p_2 < 1)$, that at least one of the outputs that produced any one neuron in $S_y$ is located in $S_x$, the probability of $\forall n(n \in S_y \wedge n \notin S_x)$ is $(1-p_2)^{|S_y|}$. Thus the probability of $S_x \cap S_y = \emptyset$ is $(1-p_1)^{|S_x|} * (1-p_2)^{|S_y|}$ (this is the probability of two branches disconnecting). This product will be very small because the values of $|S_x|$ and $|S_y|$ are somewhat large, then the probability of existing multiple maximum-connected branches will be very small. The conclusion reached is that it may be ignored in the case that the sample cannot actually produce a fully-connected representation. In other words, a large number of redundant connections exist.

Considering the second situation, in which the neurons in the test sample (*E*) are not strong enough to stimulate all neurons in the original sample (*S*). It is assumed that the original sample consists of |*S*| active neurons, and they are used as the stimulating source to produce a connected micro structure to materialize memory. A portion of neurons in *S* take the role of "seeds". A neuron is referred to as a seed if none of its input neurons are activated Otherwise, according to the modulation learning algorithm (see Section 5), whether the connection is established in the stage of free radiation or in the stage of modulation learning, a neuron is probably activated by other activated neurons, i.e. the input field of seed neurons and the fan-out closures that are formed by other active neurons transmitting activation are not connected. It's essential to determine the probability that any neuron is in its input field but is not in the fan-out closure produced by *S*-1 neurons. This is similar to 6.1, which is $1-(\frac{|S|-1}{N})^{\frac{1}{M}}$, and the mathematical expectation for the number of seed neurons is $|Seed|=|S|*(1-(\frac{|S|-1}{N})^{\frac{1}{M}})$, $0<|Seed|<|S|$. The two factors that cause the test sample not to recover a memory correctly are: that *E* cannot cover the seed set and the other is that *E* covers the seed set, but cannot activate any node within the activation loop. In common, at least three neurons are needed to form a loop, its probability is $(\frac{M}{N})^3$, as $N>>M$, and its value must be very small. Therefore, in fact, a loop composed of more than 3 nodes is almost impossible. Evidently, only the first factor must be considered in this situation, and the probability is $(\frac{|E|}{|S|})^{\frac{|Seed|}{|S|}}$ where the seed node set is fully covered by |*E*| nodes, when they spread into *S*. In the experimental system *N* (the number of nodes) is 1024 and *M* (the number of connections) is 16, the success probability is 91.01%, when |*S*|=100 and |*E*|=50; it is 88.11% when |*S*|=512 and |*E*|=200; and it reaches 96.00% when |*S*|=512 and |*E*|=200. From the above, the success probability is much higher in non-extreme conditions. The memory fails inevitably when |*E*| <|*Seed*| and the key is the uniformity of the distribution of the seed set.

From the probability analysis in this section, it is not difficult to form a reliable memory. Humans tend to recall what they most desire to remember.

6.3 The analysis of memory capacity

In this paper, the physical realization of memory is constructing a connected-structure, in which a variable number of neurons produce various connected-structures. For memory capacity, the total number of nodes in the network and the total number of connections from one node to another node are highly significant. However, the capacity is not only dependent on the number of connections and nodes in a network, but also on the resolution and the connecting way provided by the fitting function. In cases where only a natural connection is concerned, if there are *N* neuron nodes and *M* input and output connections individually for each node in a network, $C_N^1 * \prod_{i=1}^{n-1} C_M^1$ connection patterns could be produced by *n* nodes. In theory, this means that the above number is also the number of totally different memory contents, especially as the value *n* is variable, so the value of $\sum_n (C_N^1 * \prod_{i=1}^{n-1} C_M^1)$ must have a high magnitude. From this it is very clear how difficult it is to estimate the memory capacity of the human brain.

The above analysis is closely related to the massive parameters in the network. To ensure that the neural network works normally, each parameter must be set properly and precisely. In human beings these parameters have been set and adjusted as a consequence of evolution.

## 7. A Micro-Explanation for Cognitive Operation based on Dynamic Processes

Figure 6 shows an experimental result. (a) In the memorizing stage, the first sequence is a sample that needs the network to memorize. This sample is composed of 9 schemes, and each of them is a snapshot of activities of a neuron-array. (b) In the recalling stage (which proceeds after the network has remembered several samples, and the capacity of discriminating patterns of memory are observed), while a test snapshot similar to the first scheme appears, the network succeeds in recalling the second sequence. In addition to the first scheme itself, the new sequence includes 8 other subsequent schemes. It must remember precisely each scheme itself, as well as remember the sequences of the schemes, because the presentation of sample is treated as a continuous temporal sequence. Considering the summation efficiency of time mentioned in Section 4, the temporal correlations of schemes are resolved into the relations of neurons' input/output mapping and the fitting functions, and therefore, the schemes and their successive orders are smoothly remembered.

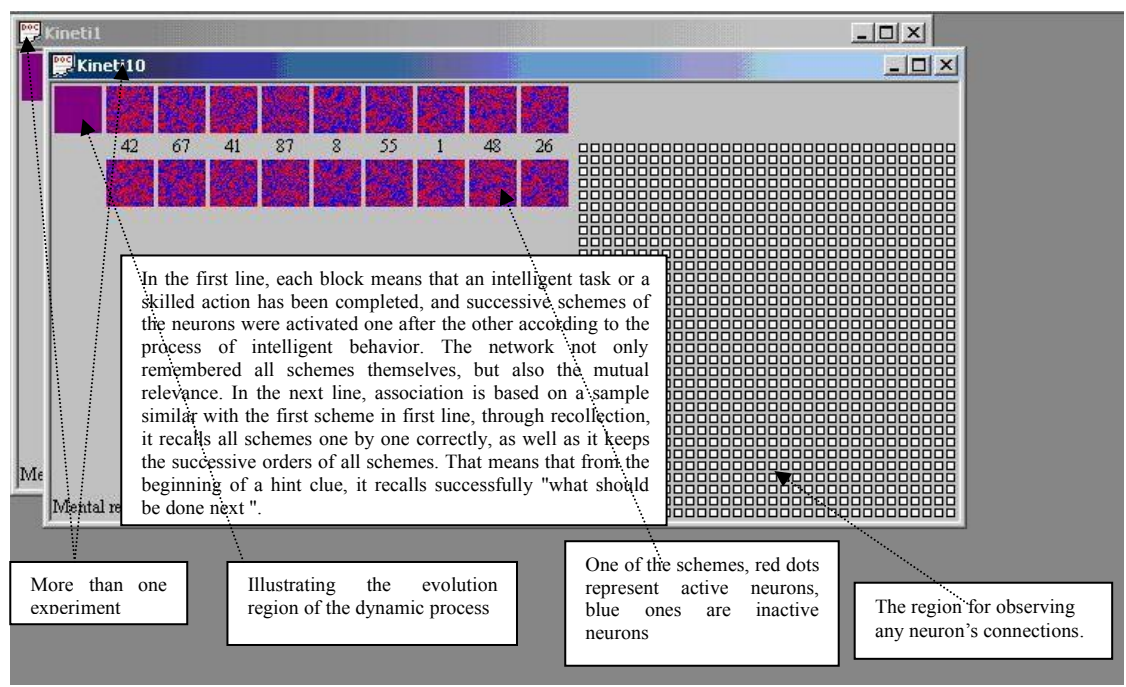

Fig. 6. Memorizing and Recalling an Intelligent Task

The first line may be considered as a recording of the activities of a group of neurons in some region of a subject's brain, when they are asked to execute a task, e.g., "push the bell once with the right hand when a red circle appears". The execution of this task is divided into multiple steps. If these steps are adequate, then any of them is regarded as doing a cognitive operation. Then each scheme is considered as the physical representation related to a certain cognitive operation. This kind of scheme sequence is very similar to the images of NMR, PET, CT, and fMRI in brain imaging research, except that differences exist in the scope of the brain's area which is observable, and in the granularity of resolution. In the same way, the first line is also seen as a driver planning a route to their destination, or as a tennis player tracking a tennis ball's trajectory and performing a series of coherent actions to hit it back, or as the usual problem-solving steps dealing with a certain problem, or as a chain of association kindled by a certain scene, or as an inference process to prove a geometric theorem. Any mental activities like the foregoing examples may be disassembled into a queue combined by multiple snapshots.

Following is a complicated example of problem-solving by means of reduction. Suppose that a doctor was facing a brand-new problem without a ready-made solution: an intracranial tumor needs to be killed by radiation without harming any healthy tissue. Following considerable thought, a physician applied the analogy reasoning method to this problem by simulating a general dividing soldiers into many groups to besiege a castle in different directions, and irradiated that tumor through multiple but power-mild beams in many different directions. Problems are soluble by being broken into many sub-problems, and the representation of each sub-problem might trigger a known solution. Due to the similar representations of two sub-problems' input or output status, such as the similar centripetally focusing modes of the two preceding problems, this similarity acts as a clue to start association and bridge two problems, and the existed problem-solving steps for an old problem could be awakened one by one to settle new problems, just like the schemes being awakened one after another in Fig. 6. Thus the neural dynamic process of problem-solving illustrated by the schemes evolving in Fig.6 is equivalent to the relaying and nesting of sub-problems' solutions. The great advantage of explaining problem-solving at the neural dynamic level is that the problem-solving methods are integrated together smoothly, no matter whether the original problems belong to the same modality field or not, because at this micro-representation level both the domain of the sub-problems and the interface of schemes are exactly the same or are not important, and it will be feasible if they are similar. For example, for the two problems above, the representations in respect to centrality are similar, so a similar scheme naturally introduced a ready-made strategy for solving the new problem.

No matter how difficult the psychological process is that deals with a task, no matter how abundantly the psychological resources and cognitive types are concerned, their neural processes explored by the approach of brain imaging are homogenous, and the basic rules governing them are essentially consistent as well. In other words, the cognitive behaviors, that have various differences with respect to psychological description, have little difference in nature when they are reduced to corresponding neural processes. These neural processes are all based on homogeneous representation without any domain boundary, i.e., any psychological process are explained in view of neural dynamic processes. This is the thinking of physicalism, in which the brain-process and psychological process are thought to be the same. The school of bionics depends on the theory of computational cognitive neuroscience, which applies the artificial neural network to directly represent the information, and assigned the kernel position to it in the cognitive structure.

Representation takes the role of bridge, joining all modules of cognitive information processing, and causing the correlation in cognition. The conscious mental activities, such as sensing, perception, attention, memory, learning and thinking, and the unconscious neural actions, such as reflex, control and modulation, all are regarded uniformly as the electrochemical movement of neurons— a kind of real material movement.

The study of memory's mechanism is substantively applicable to the representation of perception. There is a question whether the representation of perceptual status is a space-localized distribution in the array of neural network, or if it is a distribution in the temporal sequence of a network's running states as a whole. The answer to this is important. The former shows that the function of the computing unit is specialized, whereas the latter shows that the temporal phase of network dynamics is specialized, but both belong to the theory of neurons and the theory of equilibrium memory. The perception is stable and recognizable, and is aware and manipulable, so that representing perception from a neural computing perspective is able to explain the above characteristics.

The following real-world example clarifies how the above neural dynamic model applies to artificial intelligence (AI): A taxi driver drives a passenger. In this course the driver encounters a number of problems: the need to plan the driving route, interpret traffic signals and symbols, judge the status of traffic, respond to inquiries, and behave as a travel/shopping guide. All these intelligent behaviors are typical AI problems. Individually, any of them is a complex problem, which arises in a primitive context without any simplification. It is this highly complex when they arrive together, the key to which is switching problems fluently from one modality to another. The information stream involved may not be disturbed. Various capabilities are needed by the driver: (a) prior knowledge; (b) the ability to merge information from multiple sensors;(c) the ability to apply knowledge to current situation; and (d) the ability to provide a tractable solution.

If a robot is asked to act as that driver, one would find that this seemingly trivial work is beyond the ability of current technology. Each branch of AI, including knowledge representation, planning, pattern recognition, natural language understanding, inference, problem solving, searching, machine learning and robotics, has established many theories and practices individually, but it is difficult to program a robot which needs to integrate so many diverse methods from different backgrounds. This is because those existing methods were developed isolatedly in different branches, their theoretical bases, application areas, formalizations, mechanisms, the forms of knowledge and the forms of representation were not considered systematically under a cognition framework. For humans this is a trivial task, the cognitive operations underlying it generally exist, and these operations retrieve and use knowledge or experience that is stored in memory. During the procedure, knowledge is retrieved and used effectively, subtasks smoothly switch between each other, cues in different modalities appropriately serve to solve the problem, and the problem-solving strategies can be scheduled and linked correctly at a variety of levels. All of these depend on a consistent mode, in which knowledge and experience are represented in and recalled from memory in accordance with their unique meaning[12,13]. The implementation ways of the brain memorizing these meanings are consistent and identical, and the homogeneity of their circuits makes dynamics easily merge and shift.

The neural dynamics of cognitive operations allows a general framework for AI to integrate intelligent behaviors, which have a variety of scope, granularity, measure and form. Under this framework all these behaviors are modeled with the same paradigms, and are seamlessly integrated and cooperate with each other in solving more complex real problems. With such a representation basis, it is possible to investigate how deeply those major fields of AI, such as cognitive model, inference, knowledge representation, problem solving, language processing, pattern recognition, computer vision and machine learning are correlated in this general cognitive structure. Two applications of this hypothesis are:

(1) how language understanding depends on semantics and inference, and how semantics is obtained from learning, problem solving and perception, and

(2) how pattern recognition and visual interpretation relies on representation, inference and prior knowledge, and how prior knowledge relies on learning and memory.

In Ref. [14] it is stated that "human intelligence discriminates itself by the way it observes and analyzes problems with many granularities, while computers cannot represent objects at different levels". The reason is that AI overemphasizes the abstract and formal description of problem. This work does not aim to provide a concise, convenient and direct description of knowledge, yet expects to discover the hidden relationship of knowledge (also experience and skill) acquisition, representation and usage, and the indivisibility and conditionality among them. They are all part of research on representation and computation within the mind.

**References**

1. U.Sandler & L.Tsitolovsky, Fuzzy dynamics of brain activity, Fuzzy Sets and Systems, Vol.121, 2001, 237-245

2. Andreas V. M. Herz, Tim Gollisch, Christian K. Machens, Dieter Jaeger, Modeling Single-Neuron Dynamics and Computations: A balance of detail and abstraction, Science, Vol. 314(5796), 2006, 80-85

3. Alain Destexhe, Diego Contreras, Neuronal computations with stochastic network states, Science, Vol.314(5796), 2006, 85-90

4. Fox, Michael D.,Snyder, Abraham Z.,Vincent, Justin L.,Corbetta, Maurizio,Van Essen, David C.,Raichle, Marcus E, The human brain is intrinsically organized into dynamic, anticorrelated functional networks, PNAS, Vol.102(27), 2005, 9673-9678.

5. Viktor K. Jirsa, Connectivity and dynamics of neural information processing, Neuroinformatics, Vol.2 (2), 2004, 183-204

6. Dale Purves et al., Neuroscience (4th edition), Sunderland, MA, Sinauer Associates Inc,, 2008, 25-176

7. Robert Solso, Cognitive Psychology (6th edition in 2001) (in Chinese, He Hua et al. translated), Jiangsu Education Publishing House, Nanjing, 2006, 177-198

8. Yoshio Sakurai, How do cell assemblies encode information in the brain?, Neuroscience and Biobehavioral Reviews, Vol.23 ,1999, 785-796

9. Toshinori Deguchi, Keisuke Matsuno, Naohiro Ishii, On Capacity of Memory in Chaotic Neural Networks with Incremental Learning , Lecture Notes in Computer Science, Springer Berlin, Vol.5178, 2008, 919-925

10. Osada T, Adachi Y, Kimura HM, Miyashita Y, Towards understanding of the cortical network underlying associative memory, Philos Trans R Soc Lond B Biol Sci., Vol.363(1500), 2008, 2187-2199

11. Vogel DD, A neural network model of memory and higher cognitive functions, Int J Psychophysiol. Vol.55(1), 2005, 3-21

12. Longnian Lin, Remus Osan, Shy Shoham, Wenjun Jin, Wenqi Zuo, Joe Z. Tsien, Identification of network-level coding units for real-time representation of episodic experiences in the hippocampus, PNAS,

Vol.102(17), 2005, 6125-6130

13. Longnian Lin, Guifen Chen, Hui Kuang, Dong Wang, and Joe Z. Tsien, Neural encoding of the concept of nest in the mouse brain, PNAS, Vol.104 (14), 2007, 6066–6071

14. Zhang Bo, Zhang Lin, Theory of Problem Solving and its Application (in Chinese), Beijing: Qinghua University Publisher, 1990, 4-5